%Paper: hep-th/9501111
%From: kachru@string.harvard.edu (Shamit Kachru)
%Date: Tue, 24 Jan 95 14:59:16 -0500

\input harvmac.tex
\noblackbox

%%%%%%%%%%%%%%%%%%%%%%%%%%%%%%%%%%%%%%%%%%%%%%%%%%%%%%%%%%%%%%%%%%%%%%%%%%%
%Blackboard letters
%  The prehistoric version of this font is known as "msym". Many unfortunate
%  souls still have this old (and UGLY) ancestor of "msbm". Time to join the
%  modern world guys!

\font\blackboard=msbm10 \font\blackboards=msbm7
\font\blackboardss=msbm5
\newfam\black
\textfont\black=\blackboard
\scriptfont\black=\blackboards
\scriptscriptfont\black=\blackboardss
\def\blackb#1{{\fam\black\relax#1}}

%   Those truly poor slobs who have neither "msbm not "msym" fonts can
% substitute
%   the definition

%\def\blackb{\bf}

%   for the above font definitions or, if all else fails,
%   return to scratching symbols in the dirt with a sharpened stick.
%
\def\BC{{\blackb C}}

% Blackboard bold "1". Not in the AMS font set.

\def\CO{{\cal O}}
\def\H#1#2{{\rm H}^{#1}(#2)}

\lref\DGExact{J. Distler and B. Greene,
``Some Exact Results on the Superpotential from Calabi-Yau
Compactifications,'' {\it Nucl. Phys.} {\bf B309} (1988) 295.}
\lref\Kawai{T. Kawai and K. Mohri, ``Geometry of (0,2) Landau-Ginzburg
Orbifolds,'' {\it Nucl. Phys.} {\bf B425} (1994) 191, {\tt hep-th/9402148}.}
\lref\Martinec{E. Martinec,
``Criticality, Catastrophes, and Compactifications,''
in {\it Physics and Mathematics of Strings}, ed. L. Brink,
D. Friedan, and A.M. Polyakov, World Scientific, 1992.}
\lref\Vafa{C. Vafa and N. Warner, ``Catastrophes and the Classification
of Conformal Field Theories,'' {\it Phys. Lett.} {\bf 218B}
(1989) 51; B. Greene, C. Vafa, and N. Warner, ``Calabi-Yau Manifolds
and Renormalization Group Flows,'' {\it Nucl. Phys.} {\bf B324}
(1989) 371.}
\lref\SW{A. Strominger and E. Witten, ``New Manifolds for
Superstring Compactification,'' {\it Comm. Math. Phys.}
{\bf 101} (1985) 341.}
\lref\duality{A. Giveon, M. Porrati, and E. Rabinovici, ``Target
Space Duality in String Theory,'' {\it Phys. Rept.} {\bf 244} (1994)
77, {\tt hep-th/9401139}.}
\lref\DK{J. Distler and S. Kachru, ``(0,2) Landau-Ginzburg Theory,''
{\it Nucl. Phys.} {\bf B413} (1994) 213, {\tt hep-th/9309110}.}
\lref\DKtwo{J. Distler and S. Kachru, ``Singlet Couplings and (0,2)
Models,'' {\it Nucl. Phys.} {\bf B430} (1994) 13, {\tt hep-th/9406090}.}
\lref\Witten{E. Witten, ``Phases of N=2 Theories in Two Dimensions,''
{\it Nucl. Phys.} {\bf B403} (1993) 159, {\tt hep-th/9301042}.}
\lref\NewIss{E. Witten, ``New Issues in Manifolds of SU(3) Holonomy,''
{\it Nucl. Phys.} {\bf B268} (1986) 79.}
\lref\Mirror{S.T. Yau, ed., {\it Essays on Mirror Manifolds},
International Press, 1991.}
\lref\UY{K. Uhlenbeck and S.T. Yau, ``On the Existence of
Hermitian-Yang-Mills Connections in Stable Vector Bundles,''
{\it Comm. Pure. App. Math.} Vol. XXXIX (1986) S257.}
\lref\WitSilvtwo{E. Silverstein and E. Witten, to appear.}
\lref\Dixon{T. Banks, L. Dixon, D. Friedan, and E. Martinec,
``Phenomenology and Conformal Field Theory or Can String Theory Predict
the Weak Mixing Angle?'', {\it Nucl. Phys.} {\bf B299} (1988) 613.}
\lref\CHSW{P. Candelas, G. Horowitz, A. Strominger, and E. Witten,
``Vacuum Configurations for Superstrings,'' {\it Nucl. Phys.} {\bf B258}
(1985) 46.}
\lref\KW{S. Kachru and E. Witten, ``Computing the Complete Massless
Spectrum of a Landau-Ginzburg Orbifold,'' {\it Nucl. Phys.}
{\bf B407} (1993) 637, {\tt hep-th/9307038}.}
\lref\DistGr{J. Distler and B. Greene, ``Aspects of (2,0) String
Compactifications,'' {\it Nucl. Phys.} {\bf B304} (1988) 1.}
\lref\DSWW{M. Dine, N. Seiberg, X. Wen,  and E. Witten,
``Non-Perturbative Effects on the String World Sheet I,II,'' {\it
Nucl. Phys.} {\bf B278} (1986) 769, {\bf B289} (1987) 319.}
\lref\Morrison{P. Aspinwall, B. Greene, and D. Morrison, ``Calabi-Yau
Moduli Space, Mirror Manifolds, and Spacetime Topology Change in
String Theory,'' {\it Nucl. Phys.} {\bf B416} (1994) 414,
{\tt hep-th/9309097}.}
\lref\GP{B. Greene and M. Plesser, ``Duality in Calabi-Yau Moduli Space,''
{\it Nucl. Phys.} {\bf B338} (1990) 15.}
\lref\CP{P. Candelas, X. de la Ossa, P. Green, and L. Parkes, ``A Pair
of Calabi-Yau Manifolds as an Exactly Soluble Superconformal Theory,''
{\it Nucl. Phys.} {\bf B359} (1991) 21.}
\lref\models{S. Kachru, ``Some Three Generation (0,2) Calabi-Yau Models,''
Harvard preprint to appear.}
\lref\Candelas{P. Candelas, ``Yukawa Couplings Between (2,1) Forms,''
{\it Nucl. Phys.} {\bf B298} (1988) 458.}

\Title{\vbox{\hbox{HUTP--95/A001}\hbox{UTTG--02--95}
\hbox{\tt hep-th/9501111}\vskip -.25in }}
{Duality of (0,2) String Vacua}
\centerline{Jacques Distler$^\star$}
\smallskip\centerline{\it Theory Group}
\centerline{\it Department of Physics}
\centerline{\it University of Texas}
\centerline{\it Austin, TX 78712}
\medskip
\centerline{and}
\medskip
\centerline{Shamit Kachru$^\dagger$}
\smallskip\centerline{\it Lyman Laboratory of Physics}
\centerline{\it Harvard University}\centerline{\it Cambridge, MA 02138 }

{\parindent=-5pt

\footnote{}{\par
${}^\star$Email: {\tt distler@utpapa.ph.utexas.edu}\ .\par
${}^\dagger$Junior Fellow, Harvard Society of Fellows.
Email: {\tt kachru@string.harvard.edu}\ .\par
}
}

\vskip .25in

We discuss
a duality of (0,2) heterotic string vacua
which implies that certain
pairs of (0,2) Calabi-Yau
compactifications on topologically distinct target manifolds
yield identical string theories.
Some complex structure moduli in one model are interpreted in the
dual model as deforming the holomorphic structure of the  vacuum gauge bundle
(and {\it vice versa}).
A better understanding of singularity resolution for (0,2) models
may reveal that this duality of compactifications
on singular spaces is
part of a larger story, involving smooth topology-changing processes
which
interpolate between the (0,2) models on the resolved spaces.

\bigskip

\Date{\it January 1995} %replace this line by \draft  for preliminary versions

%\draft

\newsec{Introduction}

(0,2) superconformal field theories with integral $U(1)$ charges and the
appropriate central charge can be used to compactify the heterotic string
to four dimensions, yielding string models with unbroken N=1
spacetime supersymmetry \Dixon.   Unlike their
(2,2) counterparts (which manifest an
unbroken $E_{6}$ gauge symmetry), (0,2) theories can also
yield models with an effective $SU(5)$ or $SO(10)$ unification group, and
are therefore of particular interest for model
building \refs{\NewIss,\DistGr,\models}.

However, little is known about general (0,2) Calabi-Yau
theories.  In part, this is
because the issue of conformal invariance has been clouded by
worldsheet instantons \DSWW.  Recent work \refs{\DKtwo,\WitSilvtwo}\
indicates that contrary to previous expectations, (0,2) Calabi-Yau
$\sigma$-models are {\it not} destabilized by such nonperturbative
sigma model effects.\foot{The conclusions of the forthcoming paper
\WitSilvtwo\ in particular are stronger than those of \DKtwo\
 and imply that all of the models considered here are
{\it bona~fide} solutions of string theory.}
Therefore, it is important to extend our
detailed knowledge of (2,2) vacua to their (0,2)
counterparts.

One of the most interesting ``stringy'' phenomena encountered in the
exploration of (2,2) models is mirror symmetry \refs{\GP,\CP,\Mirror}\ :
The same physical theory can be obtained by considering string
propagation on a manifold $M$ or on its topologically distinct
mirror $\tilde M$.  This symmetry (like other examples of duality in
string theory) is completely unexpected from the perspective of a
Kaluza-Klein theorist.

One might wonder what other sorts of duality occur in more
general (0,2) models.  In this paper we explore a duality first
noted in \DK.
A (0,2) Calabi-Yau model requires for its specification
both a manifold $M$ and a stable, holomorphic vector bundle
$V \rightarrow M$, the vacuum gauge bundle.  The
duality we
discuss relates a (0,2) model with data $(M,V)$ to a
(0,2) model with data $(\tilde M, \tilde V)$, with $M$ and $\tilde M$
topologically distinct.  As in mirror symmetry, the dual pairs
should provide
different descriptions of the {\it same} physical theory.

An interesting possibility, also mentioned in \DK, is that this
duality is a small part of a larger story.  We will find numerous
examples of dual pairs, but in each of our examples singularities
on $M$ or $\tilde M$, which are ``frozen in'' on the part of the
moduli space we can study, will complicate the analysis.
It could be that when the resolution of these singularities is
properly understood, one will find phenomena analogous
to those of \refs{\Witten,\Morrison}.  Namely, it could be that there will be
additional moduli which resolve the singularities, and that
by varying these moduli we can pass through a ``wall'' separating
topologically distinct ``phases'' -- corresponding to the
smooth $(M,V)$ and $(\tilde M, \tilde V)$ theories.
If this is the case, the duality we see here is merely the
statement that on this wall -- in the theory
on the singular manifolds --
one can ascribe either geometrical interpretation to the
conformal theory.

The moduli of a (0,2) Calabi-Yau model come in
three types: Elements of $\H{1,1}{M}$ and $\H{2,1}{M}$ which
deform the K\"ahler and complex structure of the base manifold $M$,
and elements of $\H{1}{M,{\rm End}(V)}$ which correspond to deformations
of the holomorphic structure of the gauge bundle $V$.
The duality we discuss exchanges some elements of $\H{2,1}{M}$
with elements of $\H{1}{\tilde M, {\rm End} (\tilde V)}$.

In \S2 we review the phases picture of the moduli spaces of
(0,2) Calabi-Yau models \refs{\Witten,\DK}, and provide
evidence for our (0,2) duality by finding
examples of distinct (0,2) Calabi-Yau
models with isomorphic Landau-Ginzburg phases. This extends in a more
or less straightforward manner to an argument that the models
remain isomorphic on the full subspace of their moduli spaces
described by our $U(1)$ gauged linear sigma models.
\S3 contains
a large radius check of the equivalence of pairs related by
the phenomenon of \S2.  We make the easy
observation that the unnormalized Yukawa couplings of the
generations in the dual descriptions are equal, and also make some
remarks about the dependence of the Yukawa couplings
in such (0,2) models on the K\"ahler modulus.
Some open questions are discussed in \S4.

\newsec{Phase Diagrams and Dual Pairs}

A useful tool in understanding the
moduli spaces of many Calabi-Yau models is the gauged linear sigma
model, introduced in this
context by Witten \Witten\ (an equivalent description, in the language of
toric geometry, is given in \Morrison ).
The basic idea is to study spaces of gauged linear sigma models which, in
the infrared limit, approach the conformally invariant theories of interest.
Following \DK\ (and in particular using the same notation
and (0,2) multiplets explained in that paper), we now show
the existence of dual pairs of (0,2) vacua by studying appropriate
gauged (0,2) linear sigma models.

\subsec{A Class of (0,2) Calabi-Yau Models}

The data involved in specifying
the (0,2) theories of interest to us includes a
Calabi-Yau manifold $M$ and a stable, holomorphic vector bundle
$V\rightarrow M$ satisfying
\eqn\constr{c_{2}(V) = c_{2}(TM),~~~~c_{1}(V) = 0~{\rm mod} ~2~.}
The first condition is the familiar anomaly cancellation condition,
while the second condition guarantees that the bundle $V$ admits
spinors.

Beyond these topological conditions, there are of course perturbative
requirements for conformal invariance of a (0,2) Calabi-Yau
sigma model \NewIss.
At lowest order, we must choose the K\"ahler metric $g_{i\bar j}$
on $M$ to be the Ricci-flat metric whose existence is guaranteed by
Yau's theorem, and in addition we must choose the connection on $V$
to satisfy the Donaldson-Uhlenbeck-Yau equation
\eqn\duy{g^{i\bar j}F_{i\bar j}~=~0.}
The integrability condition for existence of a solution to \duy\ is
\eqn\integ{\int_{M} J\wedge J\wedge c_{1}(V) ~ = ~0}
where $J$ is the K\"ahler form of $M$ \UY.  We
will meet this condition
by requiring $c_{1}(V) = 0$.  Higher orders of sigma model
perturbation theory do not lead to any new conditions on $M$
or $V_{1,2}$.

We will concentrate on complete intersection Calabi-Yau manifolds
defined by the intersection of N hypersurfaces
\eqn\manif{W_{i}(\phi)~=~0,~i=1,\dots,N~}
of degree $d_{i}$ in some $WCP^{3+N}_{w_{1},\cdots,w_{4+N}}$
with homogeneous coordinates $\phi_{1},\cdots,\phi_{4+N}$.
We use bundles $V$ defined as the kernel of an exact sequence
\eqn\bundle{0 \to V\to \bigoplus_{a=1}^{\tilde r+1} \CO(n_{a})
{\buildrel {\otimes F_{a}(\phi)} \over {\hbox to 30pt{\rightarrowfill}}}
\CO(m) \to 0}
where $m$ and $n_{a}$ are positive integers,
$\CO(j)$ denotes the $j$th power of the hyperplane bundle of the
ambient weighted projective space, and the $F_{a}(\phi)$ are polynomials
homogeneous of degree $m-n_{a}$ in the $\phi$s.
The constraints \constr\ are translated to conditions on the integers
$m$, $n_{a}$, and $d_{i}$:
\eqn\constag{m^{2}-\sum_{a=1}^{\tilde r +1} n_{a}^{2} =
\sum_{i=1}^{N} d_{i}^{2} - \sum_{j=1}^{4+N} w_{j}^{2},~~~~
m = \sum_{a=1}^{
\tilde r +1} n_{a}~.}
In addition, one of course can write the Calabi-Yau condition as
\eqn\cy{ \sum_{i=1}^{N} d_{i} = \sum_{j=1}^{N+4} w_{j}~.}

\subsec{A CY Manifold as a Gauged Linear Sigma Model}

Let us now rephrase \S2.1 in the language of quantum field theory.
As an example, let us start by discussing the gauged
linear sigma model description of the
complete intersection $M$ of three hypersurfaces $W_{i}(\phi)=0$
of degrees $3,6,6$ in $WCP^{6}_{1,1,1,1,3,3,5}$
with gauge bundle $V$ specified by $m=7$
and $\{ n_{a}\} =  \{ 1,1,2,3\}$.
Introduce a worldsheet $U(1)$ gauge field, seven chiral superfields
$\Phi_{j}$, four left-moving fermi multiplets $\Lambda_{a}$,
three left-moving fermi multiplets $\Sigma_{i}$,
and one additional chiral
superfield $P$.  Assign the $\Phi_{j}$ gauge charges
$w_{j}=1,1,1,1,3,3,5$, the
$\Lambda_{a}$ charges $n_{a} = 1,1,2,3$, the $\Sigma_{i}$
charges $d_{i}=-3,-6,-6$, and $P$ charge $m=-7$.
In addition to the normal kinetic terms for all of these multiplets,
introduce a $U(1)$ Fayet-Iliopoulos D-term with coefficient $r$
and a (0,2) superpotential
\eqn\superpot{{\cal W}
 = \int d^{2}z ~d\theta~ \Sigma_{i}W_{i}(\phi) +
P \Lambda_{a} F_{a}(\phi)~.}

Integrating out the D auxiliary field in the gauge multiplet and the
auxiliary fields in the Fermi multiplets, we find the scalar potential
\eqn\scalarpot{U~=~\sum_{i=1}^{3} \vert W_{i}(\phi) \vert ^{2} +
\vert p \vert^{2} \sum_{a=1}^{4} \vert F_{a}(\phi) \vert^{2}
+ {e^{2}\over 2} \left( \sum_{j=1}^{M+4} w_{j}\vert \phi_{j} \vert^{2}
- m\vert p \vert^{2} - r \right) ^{2}~.}

Now, we study the infrared behavior of this theory by focusing on the
locus of vanishing $U$.  While in general this is not necessarily a
good approximation to the infrared physics, for $\vert r\vert$ very large
studying only the ground states and the massless excitations around
them does suffice.  Integrating out the massive fields, for large
$\vert r\vert$,
simply induces corrections to the parameters of the low energy theory.

For $r$ very large and positive, we see that the minimum of
\scalarpot\ is obtained when
\eqn\scalmin{p=0,~~
\sum_{j} w_{j}\vert \phi_{j}\vert^{2}
= r,~~ W_{i}(\phi)=0~.}
Dividing the space of solutions of
$\sum_{j} w_{j}\vert\phi_{j}\vert^{2}$ by the $U(1)$ gauge group
precisely gives us $WCP^{6}_{1,1,1,1,3,3,5}$ with K\"ahler class
proportional to $r$.  The last constraint in \scalmin\
then tells us that
the space of ground states for large $r$ is precisely the
Calabi-Yau manifold $M$.

What about the fermions?  As in the (2,2) theories discussed in
\Witten, the superpartners of the $\phi_{i}$ (which are right-moving
fermions) transform as sections of the tangent bundle $TM$.
The Yukawa coupling
\eqn\yuk{\psi_{p}\lambda_{a}F_{a}(\phi)}
with the fermionic partner of $p$
gives a mass to one linear combination of the $\lambda$s, while the
rest transform as sections of the rank $\tilde r$ bundle $V$.
Therefore, for large positive $r$ we have recovered the (0,2)
Calabi-Yau sigma model
described by $V\rightarrow M$.

Next, we see what happens for $r$ very negative, where semiclassical
reasoning again should be a good guide to the infrared physics.
Minimizing \scalarpot\ we this time find
\eqn\smintwo{\vert p \vert^{2} = r,~~\phi_{j}=0~.}
The gauge symmetry is broken to a $Z_{7}$ because $p$ has gauge charge
$-7$; and $p$ and $\psi_{p}$ become massive and drop out of the theory.
What remains is then a (0,2) Landau-Ginzburg theory, with
superpotential (after rescaling fields to absorb the VEV of $p$)
\eqn\lgpot{\int d^{2}z~d^{2}\theta~\Sigma_{i}W_{i}(\phi) +
\Lambda_{a}F_{a}(\phi)~.}
The $Z_{7}$ discrete gauge symmetry means that we wish to study not
the theory \lgpot,\ but its $Z_{7}$ orbifold.
So by studying the ``phases'' of the (0,2) gauged LSM, we have been
able to recover the Calabi-Yau/Landau-Ginzburg correspondence
\refs{\Martinec,\Vafa}.

Note one interesting fact about the superpotential \lgpot\ (and
indeed the full effective Lagrangian in the Landau-Ginzburg phase) --
the model has ``forgotten'' about the distinction between the
defining equations of the
manifold $W_{i}$ and the data $F_{a}$ defining the
holomorphic structure of the gauge bundle.  While they enter in
the Lagrangian in different ways at large radius, here at
small radius they are on an equal footing.

\subsec{Another CY Manifold as a Gauged Linear Sigma Model}

Now, lets repeat the story of \S2.2, this time on the manifold
$\tilde M$ defined by the intersection of degree $4,5,6$
hypersurfaces
\eqn\tildeM{\tilde W_{i}(\phi) = 0}
again in $WCP^{6}_{1,1,1,1,3,3,5}$.
This time, we choose $\tilde V$ to be defined by
$m=7$ and $\{ n_{a} \} = \{ 1,1,1,4 \}$.

At large positive $r$ we recover once again the (0,2) CY sigma model
with target $\tilde M$, with the right-moving fermions transforming
as sections of $T\tilde M$ and the left-movers as sections of
$\tilde V$.  This sigma model is {\it a~priori} unrelated to the one
described in \S2.2 -- in fact, from the viewpoint of classical geometry
the pairs $(M,V)$ and $(\tilde M, \tilde V)$ are manifestly distinct.
For example, $M$ and $\tilde M$ have distinct orbifold Euler characters
\eqn\char{\chi(M) = -216,~~\chi(\tilde M) = -192~}
(in calculating \char\ one has to resolve a $Z_{5}$ fixed point on
$M$ and two $Z_{3}$ fixed points on $\tilde M$).

Let's proceed to take the gauged
linear sigma model description of the theory
with target $\tilde M$ and vacuum gauge bundle $\tilde V$
down to $r \rightarrow -\infty$.  Having introduced gauge
charges and such as in \S2.2, we see that at $r\rightarrow -\infty$
we are left with a
Landau-Ginzburg theory possessing a discrete $Z_{7}$ gauge
symmetry with superpotential
\eqn\supertwo{\int d^{2}z~d^{2}\theta~\tilde \Sigma_{i}\tilde W_{i}(\phi)
+ \tilde \Lambda_{a}\tilde F_{a}(\phi)~.}
Now we notice something interesting.
The polynomials $\tilde W_{i}(\phi)$
have degrees $4,5,6$ and the polynomials
$\tilde F_{a}(\phi)$ have degrees $6,6,6,3$.  Looking back to \S2.2,
in the Landau-Ginzburg phase of $(M,V)$ the polynomials $W_{i}(\phi)$
have degrees $3,6,6$ while the polynomials $F_{a}(\phi)$ have degrees
$6,6,5,4$.

Now suppose we choose the data defining the two models as follows:
\eqn\map{W_{1}(\phi)=\tilde F_{4}(\phi),~~W_{2}(\phi)=\tilde F_{3}(\phi),
{}~~W_{3}(\phi)=\tilde W_{3}(\phi)~}
\eqn\maptwo{F_{1}(\phi)=\tilde F_{1}(\phi),~~F_{2}(\phi)=\tilde F_{2}(\phi),
{}~~F_{3}(\phi)=\tilde W_{2}(\phi),~~F_{4}=\tilde W_{1}(\phi)~~.}
Then the full Lagrangians defining the Landau-Ginzburg
``phases'' of the two models are identical, with the simple change
of notation
\eqn\mapthree{(\Sigma_{1},\Sigma_{2},\Sigma_{3},\Lambda_{1},\Lambda_{2},
\Lambda_{3},\Lambda_{4}) \rightarrow (\tilde \Lambda_{4},\tilde \Lambda_{3},
\tilde \Sigma_{3},\tilde \Lambda_{1},\tilde \Lambda_{2},\tilde \Sigma_{2},
\tilde \Sigma_{1})~.}
So at the Landau-Ginzburg radius there is a duality which exchanges
the two models, with the ``duality map'' given in \map,\maptwo,\mapthree.

This proves that the two naively distinct Calabi-Yau sigma models
become isomorphic in their Landau-Ginzburg phases.  Furthermore,
the K\"ahler modulus $r$ representing the overall size of the ambient
$WCP^{6}$ arises as a twist field in the Landau-Ginzburg theory.
Perturbation theory in this twist field will be identical for the two
theories, so the identity between them extends off the Landau-Ginzburg
locus to an open set in their K\"ahler
moduli spaces (the region of convergence
of the perturbation expansion in the twist field).
But two models which agree in such a region in the (complexified) K\"ahler
moduli space spanned by $r$ and the worldsheet $\theta$ angle must
agree on the whole $(r,\theta)$ plane.  So we find that the two models
remain isomorphic for all values of $r$ (and $\theta$).

It is important to note that in the discussion of the K\"ahler moduli
space above, we really mean only the K\"ahler modulus corresponding
to the parameters $(r,\theta)$ in the linear sigma model.  In the (2,2)
models on these manifolds, the full K\"ahler moduli spaces would be
three (complex) dimensional, with two additional elements of
$h_{1,1}$ being introduced by resolving the $Z_{5}$ fixed point
on $M$ and the two $Z_{3}$ fixed points on $\tilde M$.  In these
(0,2) models, the status of these additional elements of $h_{1,1}$
is less clear.   It could be that once these moduli, and perhaps other
additional moduli which are related to the resolution of the
singularities of $(M,V)$ and $(\tilde M, \tilde V)$, are added,
the duality of this subspace of the moduli spaces of $(M,V)$ and
$(\tilde M, \tilde V)$ will be seen as part of a larger story
involving topology change as in \refs{\Witten,\Morrison}.

We should also mention now that the isomorphism \map\maptwo\mapthree\
manifestly mixes the $W_{i}$ and $\Sigma_{i}$ with the
$\tilde F_{a}$ and the $\tilde \Lambda_{a}$.  But
perturbations of the $\tilde F_{a}$ correspond, at large radius, to changes
in the holomorphic structure of $\tilde V$, while perturbations of
the $W_{i}$ correspond to changes in the complex structure of $M$.
So we see that the duality we are studying involves the exchange
of elements of $\H{2,1}{M}$ with elements of $\H{1}{\tilde M,
{\rm End} (\tilde V)}$.

\subsec{More Examples}

In \S2.3 we have seen the game that one has to play in order to find
(0,2) models which will be dual pairs.
Take two complete intersection manifolds in the same weighted projective
space and choose vacuum gauge bundles over each with
the same $m$.  Then
as long as the full sets of
degrees of the $W$s and the $F$s in one model and the $\tilde W$s and
$\tilde F$s in the other model coincide, the two theories
will be isomorphic
in their Landau-Ginzburg phases.  So by searching for different
complete intersections and bundles in the same projective space
which satisfy these criteria, we can find dual pairs.

In Table 1 below, we list several examples.  Under $M$ and $\tilde M$
we list the degrees of the hypersurfaces defining the complete
intersection manifolds, under $V$ and $\tilde V$ we give the data
$(m;\{n_{a}\})$, and under $\{w_{j}\}$ we list the weights
of the ambient weighted projective space.
In calculating the Euler characters of the singular manifolds below,
we have used the orbifold Euler characteristic formula of the second
reference in \Vafa.
The first example
is the one discussed in \S2.2-2.3 and the last example
was given in \DK.

\def\tablerule{\omit&\multispan{14}{\tabskip=0pt\hrulefill}&\cr}
$$%\ifx\answ\bigans\else\hskip-.5in\fi
\vbox{\offinterlineskip\tabskip=0pt\halign{\hskip -.3in
$#$\ &\vrule #&\ \ \hfil $\strut #$\hfil\ \ &\vrule #&
\ \ \hfil $#$\hfil\ \
&\vrule # &\ \ \hfil $#$\hfil\ \ &\vrule #&\ \ \hfil $#$\hfil\ \
&\vrule #&\ \ \hfil
$#$\hfil\ \ &\vrule#&\ \ \hfil $#$\hfil\ \ &\vrule#
&\ \ \hfil $#$\hfil\ \ &\vrule #\cr
&\omit&{M}&\omit&{V}
&\omit&\tilde M&\omit&\tilde V&\omit&\chi(M)&\omit&{\chi(\tilde M)}
&\omit&\{w_{j}\}&\omit\cr
\tablerule
&&3,6,6&&(7;1,1,2,3)&&4,5,6&&(7;1,1,1,4)&&-216&&-192&&1,1,1,1,3,3,5&\cr
\tablerule
&&3,11&&(12;1,1,3,7)&&5,9&&(12;1,1,1,9)&&-280&&-216&&1,1,1,3,3,5&\cr
\tablerule
&&4,4,4&&(6;1,1,1,3)&&3,4,5&&(6;1,1,2,2)&&-120&&-132&&1,1,1,2,2,2,3&\cr
\tablerule
&&3,4,7&&(8;1,1,2,4)&&4,4,6&&(8;1,1,1,5)&&-132&&-112&&1,1,2,2,2,3,3&\cr
\tablerule
&&6,6&&(8;1,1,3,3)&&5,7&&(8;1,2,2,3)&&-120&&-120&&1,1,2,2,3,3&\cr
\tablerule
&&3,8&&(9;1,1,2,5)&&4,7&&(9;1,1,1,6)&&-168&&-132&&1,1,2,2,3&\cr
\tablerule
&&5,8&&(10;1,1,2,6)&&4,9&&(10;1,2,2,5)&&-104&&-108&&1,2,2,2,3,3&\cr
\tablerule
\noalign{\bigskip}
\noalign{\narrower\noindent{\bf Table 1:}
Some examples of dual pairs of (0,2) Calabi-Yau $\sigma$-models on
Calabi-Yau threefolds.
}
}}$$

Each of these pairs, when represented as gauged linear sigma models
with a single $U(1)$ gauge group, has the feature that they
are manifestly isomorphic in their Landau-Ginzburg phases.
By the perturbative argument given in \S2.3,
we expect these theories to remain isomorphic at finite $(r,\theta)$
as well.
We have no insights about the
possibility of resolving
these models while preserving the conditions \constr.
Note however that the manifold $M$ in the fifth dual pair
listed in Table 1 is already nonsingular, while its dual is not.

\newsec{A Large Radius Test: Yukawa Couplings}

We have given general arguments that the pairs
of sigma models on topologically distinct Calabi-Yau manifolds
listed in Table 1 give isomorphic superstring vacua.
Still, this seems fairly
mysterious at large radius.  In order to convince the reader that it
is nonetheless true, we now discuss the large radius Yukawa couplings
of the first dual pair listed in Table 1.  Everything we say is
dependent on the (very plausible) assumption
that the point singularities
of $M$, $\tilde M$ do not affect our considerations, and generalizes
in an obvious way to the other examples of \S2.4.

These models both have an observable $E_{6}$ gauge group in spacetime.
Choose the defining data to obey the constraints \map\maptwo.
Computing the spectrum of the model at the Landau-Ginzburg point
following \refs{\KW,\DK}, we find 123 $\bf 27$s of $E_{6}$ and 1 $\bf \overline
{27}$ of $E_{6}$.  122 of the $\bf 27$s come from the untwisted sector and have
a deformation-theoretic representation as seventh degree polynomials
\eqn\defthe{\{P_{7}(\phi)\}~ {\rm mod}~ \{ W_{i}(\phi), F_{a}(\phi) \}~.}
This is the
expected large radius answer for
both models.  $\bf 27$s of $E_{6}$ should be
in 1-1 correspondence with elements of $\H{1}{M,V}$ and $\H{1}{\tilde M,
\tilde V}$ in the two models,
and it follows
from the sequence \bundle\ that $\H{1}{M,V}$ and $\H{1}{\tilde M,
\tilde V}$ should have such a deformation theoretic representation.

What about the extra $\bf 27$ and $\bf \overline {27}$
in the Landau-Ginzburg theory?
They both come from twisted sectors.  There are two possibilities --
either they survive at large radius and can be
understood as coming from the resolution of the
singularities, or they pair up as one leaves the Landau-Ginzburg point
(the latter possibility is consistent with all of the symmetries of the
Landau-Ginzburg theory).  We shall not try to distinguish between
these two possibilities here, though it certainly is possible to study
the relevant correlation functions in the Landau-Ginzburg theory.
Instead, we shall focus henceforth on the
122 deformation theoretic $\bf 27$s in each model, whose origin we
easily understand both at small and large radius.

At large radius, the
Yukawa couplings are given by the intersection forms
\eqn\intone{\H{1}{M,V}\otimes \H{1}{M,V}\otimes \H{1}{M,V} \rightarrow \BC}
\eqn\inttwo{\H{1}{\tilde M, \tilde V} \otimes \H{1}{\tilde M, \tilde V}
\otimes \H{1}{\tilde M, \tilde V} \rightarrow \BC}
where we use $\H{3}{M,\bigwedge^{3}V} =\H{3}{M,\CO}= \H{0,3}{M}=\BC$ and
$\H{3}{\tilde M, \bigwedge^{3}\tilde
V} = \H{0,3}{\tilde M}= \BC$. This follows from the fact that
$c_1(V)=c_1(\tilde V)=0$, so that for $V$, $\tilde V$ of rank 3,
$\bigwedge^3V$ and $\bigwedge^3 \tilde V$ are trivial.
Following the discussion in \SW\ and especially
\S4 and \S5 of \Candelas, it is easy to see what
the relevant formula for the unnormalized Yukawa couplings will be.

Consider the coupling $\kappa_{ijk}$ of three generations
represented by cohomology classes $A_{i}^{a}$, $A_{j}^{b}$ and $A_{k}^{c}$
where $A_{i}^{a} =A_{\bar m i}^{a}~d\bar z^{m} \in
H_{\bar \partial}^{(0,1)}(M,V)$.
This has an expression as an integral
over the manifold
\eqn\yukint{\kappa_{ijk} ~=~
\int_{M} \Omega \wedge A_{i}^{a} \wedge A_{j}^{b} \wedge A_{k}^{c}
{}~\epsilon_{abc}~}
where $\Omega$ is the holomorphic three-form of $M$.
Now, define
\eqn\barw{\bar \omega ~=~A_{i}^{a}\wedge A_{j}^{b}\wedge A_{k}^{c}
{}~\epsilon_{abc}~.}
It follows from \barw\ that $\bar \omega$
is a closed (0,3) form.
But on a Calabi-Yau manifold $\bar \Omega$ is the  unique element of
$\H{0,3}{M}$, so we may decompose
\eqn\bardec{\bar \omega ~=~ \kappa \bar \Omega + \bar\partial (\cdots)}
and since the exact part will not contribute in the integral \yukint,
the constant $\kappa$ in \bardec\ is
equal to the Yukawa coupling.

For the models of interest the elements $A_{i}^{a} \in H_{\bar \partial}
^{(0,1)}(M,V)$ have a deformation theoretic representation \defthe.
Similarly, the antiholomorphic (0,3) form $\bar \Omega$ can be represented
(up to scale) by the single $21^{st}$ degree polynomial not in the ideal
generated by the $W_{i}$ and the $F_{a}$.
Call the single element in the quotient
$\hat P_{21}$
%\foot{
%The reader may have expected $\hat P_{21}$ to correspond
%to the single element of $H_{\bar\partial}^{(0,3)}(M,\Lambda^{3}V)$
%and indeed it does -- but $\Lambda^{3}V$ is trivial and since $\bar
%\Omega$ is the only harmonic (0,3) form, $\hat P_{21} ~\sim~
% \bar\Omega$.}
\eqn\baromega{\hat P_{21} \in \{~\{P_{21}(\phi)\} ~{\rm mod}~
\{ W_{i}(\phi), F_{a}(\phi) \}~\}~.}
The rule for computing the (unnormalized) Yukawa couplings follows
in a straightforward manner -- if $A_{i}^{a}$, $A_{j}^{b}$ and
$A_{k}^{c}$ have polynomial representatives $P_{i,j,k}$, then
the coupling $\kappa_{ijk}$
follows from
\eqn\polycoup{P_{i} P_{j} P_{k}~ = ~\kappa_{ijk}~\hat P_{21}~
+ ~{\cdots}~.}

Now it is clear that the
unnormalized Yukawa couplings of the deformation
theoretic generations for all of the dual pairs listed in \S2.4\ will
agree.  The point is that the polynomial
representatives for the generations \defthe\ and the
three-form \baromega\ which
determine the Yukawa couplings via \polycoup\ can be chosen
to be {\it identical} for the dual pairs
precisely when
the conditions analogous to \map\maptwo\ are satisfied.

It worth emphasizing that the Yukawa couplings of the
deformation theoretic generations can also be considered directly at
the Landau-Ginzburg radius \Kawai.  The results are in fact exactly the
same as the large radius results we have just described.
This should be taken as strong evidence that the Yukawa couplings
of the deformation theoretic generations are in fact {\it independent} of
$r$ -- like the couplings between $(2,1)$ forms in (2,2)
Calabi-Yau models \DGExact.

\newsec{Conclusion}

In this paper we have described a new kind of duality of Calabi-Yau models.
Mirror symmetry exchanges deformations of the size of one manifold with
deformations of the shape of its mirror.  The duality described here
exchanges ``gravitational'' moduli of one classical solution
(deformations of the shape of the compactification manifold)
with ``gauge'' moduli
of its dual (deformations of the gauge field VEV).  Like other dualities
of string vacua \duality, this
is certainly not something our particle theory
intuition would lead us to expect.

There are many aspects of the (0,2) duality discussed here that we
do not understand, however.  Even in the examples presented, there
are mysteries.
This was clear in \S3 -- our large radius understanding of the spectrum
and interactions of charged
particles is incomplete, partially
because of difficulties with singularities.
The questions we
need to answer are:
\item{1)} When can singular (0,2)
models be resolved maintaining \constr\ ?
\item{2)} What effect
do the added moduli have on our picture of the moduli spaces?

It could well be that once singularity resolution is understood,
this duality will be seen as a phenomenon on the ``wall'' of
smooth topology changing processes connecting $(M,V)$ to $(\tilde M,
\tilde V)$.

\bigskip
\centerline{\bf{Acknowledgements}}

We are grateful to the authors of \WitSilvtwo\ for informing us of their
work prior to publication
and to P. Aspinwall for correspondence about
singularities.  The research of J.D.~is supported by NSF grant PHY-9009850,
the Robert A.~Welch Foundation and an
Alfred P.~Sloan Foundation Fellowship.  The research of
S.K. is supported by a fellowship from the Harvard Society of
Fellows and by the William F. Milton Fund of Harvard University.

\listrefs
\end